\documentclass{andp2012}
\usepackage[T1]{fontenc}
\usepackage[latin9]{inputenc}
\usepackage{textcomp}
\usepackage{amsmath}
\usepackage{amssymb}
\usepackage{graphicx}
\usepackage{epstopdf}
\usepackage{color}

\makeatletter

\newcommand{\lyxmathsym}[1]{\ifmmode\begingroup\def\b@ld{bold}
  \text{\ifx\math@version\b@ld\bfseries\fi#1}\endgroup\else#1\fi}

\@ifundefined{textcolor}{}
{%
 \definecolor{BLACK}{gray}{0}
 \definecolor{WHITE}{gray}{1}
 \definecolor{RED}{rgb}{1,0,0}
 \definecolor{GREEN}{rgb}{0,1,0}
 \definecolor{BLUE}{rgb}{0,0,1}
 \definecolor{CYAN}{cmyk}{1,0,0,0}
 \definecolor{MAGENTA}{cmyk}{0,1,0,0}
 \definecolor{YELLOW}{cmyk}{0,0,1,0}
}

\makeatother

\begin{document}
\title{Hyperbolic Dispersion in Chiral Molecules}
\author{Jie-Xing Zhao\inst{1,3}, Jing-Jing Cheng\inst{1,3}, Yin-Qi Chu\inst{1,3}}
\author{Yan-Xiang Wang\inst{1}}
\author{Fu-Guo Deng\inst{1,2}}
\author{Qing Ai\inst{1}\footnote{Corresponding author\quad E-mail:~\textsf{aiqing@bnu.edu.cn}}}
\address[1]{Department of Physics, Applied Optics Beijing Area Major Laboratory,
Beijing Normal University, Beijing 100875, China}
\address[2]{NAAM-Research Group, Department of Mathematics, Faculty of Science, King Abdulaziz University, Jeddah 21589, Saudi Arabia }
\address[3]{These authors contribute equally to this work. }
\begin{abstract}
We theoretically investigate the intra-band transitions in M\"{o}bius molecules. Due to the weak magnetic response, the relative permittivity is significantly modified by the presence of the medium while the relative permeability is not. We show that there is hyperbolic dispersion relation induced by the intra-band transitions because one of the eigen-values of permittivity possesses a different sign from the other two, while all three eigen-values of permeability are positive. We further demonstrate that the bandwidth of negative refraction is 0.1952~eV for the $H$-polarized incident light, which is broader than the ones for inter-band transitions by 3 orders of magnitude. Moreover, the frequency domain has been shifted from ultra-violet to visible domain. Although there is negative refraction for the $E$-polarized incident light, the bandwidth is much narrower and depends on the incident angle.

\end{abstract}
\maketitle

\section{introduction}

Since it was theoretically proposed in 1968 \cite{Veselago1968},
negative refraction has attracted broad interest because there is wide application for negative-index metamaterials \cite{Pendry2000,Smith2000,Bliokh2008,Khorasaninejad2017,Minovich2015,Zhao2016,Fang2016,Felbacq2017,Jia2018,Ai2018,Cheng2019}, such as achieving electromagnetic-field cloaking \cite{Leonhardt2006,Pendry2006},
facilitating sub-wavelength imaging \cite{Pendry2000}, and assisting crime scene investigation \cite{Shen2016}. Various ways have been proposed to realize negative refraction. One of them is double-negative metamaterial, in which both the permittivity $\varepsilon$ and permeability $\mu$ are simultaneously negative at the same frequencies \cite{Veselago1968,Pendry2000,Shen2014}.
However, it is difficult to realize negative permeability since the magnetic response is generally weaker than the electric response orders of magnitude. Therein it is the bandwidth of negative $\mu$ that determines the bandwidth of negative refraction. Another possible way to negative refraction is realizing hyperbolic dispersion relation \cite{Fisher1969,Smith2003,Poddubny2013,Jahani2016,Ai2018}. When two of the eigen-values of permittivity tensor are opposite in sign, the material can realize negative refraction if all the eigen-values of permeability tensor are positive \cite{Guan2017,Sreekanth2013}. In this case, the bandwidth of negative refraction is equal to the bandwidth of negative eigen-value of $\varepsilon$. Since the electric response is much stronger, the bandwidth of negative refraction of hyperbolic metamaterial is much larger than that of double-negative metamaterial.

In this paper, we investigate the possibility of realizing hyperbolic dispersion in a novel kind of chiral molecules--M\"{o}bius molecules \cite{Heilbronner1964,Walba1993}. M\"{o}bius molecules owns novel topological structures in which one can move from one side to the other side without crossing the border \cite{Ajami2003,Yoneda2014}. Previously, M\"{o}bius molecules have been suggested for metamaterials \cite{Chang2010,Fang2016,Poddar2014}, quantum devices \cite{Balzani2008,Yamashiroa2004,Zhao2009}, dual-mode resonators and bandpass filters \cite{Pond2000}, topological insulators \cite{Guo2009}, molecular knots and engines \cite{Lukin2005}, and artificial light harvesting \cite{Xu2018,Lambert2013}.
However, because it is double-negative metamaterial, the bandwidth of negative refraction in M\"{o}bius molecules is so small, e.g., $4\sim80~\mu$eV \cite{Fang2016,Cheng2019}, that it might be difficult to observe. Furthermore, since it is induced by the inter-band transitions,
the negative refraction is centered at ultraviolet frequency domain.
In this paper, we consider the hyperbolic dispersion induced by the intra-band transitions.
Because the magnetic response is reduced by a factor of $(N/\pi)^2$, the magnetic responses in intra-band transitions decrease by one order of magnitude for $N=12$ and thus the eigen-values of permeability is always positive around the intra-band transitions. Furthermore,
because M\"{o}bius molecules are chiral, the permittivity tensor is anisotropic and thus two of  its eigen-values can possess different signs. Thus, by using intra-band transitions, we can realize hyperbolic dispersion in the visible frequency domain.

\section{Permittivity and Permeability in M\"{o}bius Medium}

\begin{figure}
\begin{centering}
\includegraphics[width=9cm]{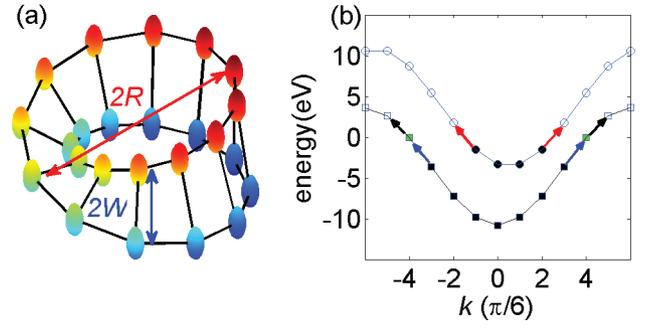}
\par\end{centering}
\caption{(a) A double-ring M\"{o}bius molecule with $2N$ atoms
and $N=12$. (b) The energy spectrum of M\"{o}bius molecule,
the black points represent the states
which are filled with two electrons, the hollow points represent
the states filled no electrons, and the green points represent
the states filled by only one electron. The transitions denoted by arrows can take place from the initial states, which are occupied by one or two electrons, to the final states, which are not occupied by two electrons. The transitions with the same transition frequency are marked with the same color. \label{Figure 1}}
\end{figure}

In this paper, we consider a general double-ring M\"{o}bius molecule
which is composed of $2N$ atoms as shown in Fig.~\ref{Figure 1}(a).
Here $W$ and $R$ are the radius of the carbon atom and the M\"{o}bius ring respectively.
$2W$ denotes the width of the M\"{o}bius ring. The two sub-rings of the M\"{o}bius molecule are linked end to end.

We consider the M\"{o}bius ring as the conjugated molecule and thus we
can use H\"{u}ckel molecular orbital method to describe the coherent dynamics in the ring. Because all the atoms of M\"{o}bius ring are of the same species, the
site energy difference between the two sub-rings $\epsilon$ vanishes. Thus the Hamiltonian for the single electron of the system can be written as \cite{Zhao2009}
\begin{equation}
H=\sum_{j=0}^{N-1}\left[A_{j}^{\dagger}MA_{j}-\xi\left(A_{j}^{\dagger}A_{j+1}+\mathrm{h.c.}\right)\right],
\end{equation}
where
\begin{align}
A_{j} & =\left[\begin{array}{c}
a_{j}\\
b_{j}
\end{array}\right],\\
M & =\left[\begin{array}{cc}
0 & -V\\
-V & 0
\end{array}\right],
\end{align}
$a_{j}$ ($b_{j}$) is the annihilation operator at the $j$th site
of sub-ring a (b), $V$ ($\xi$) is the inter-sub-ring (intra-sub-ring)
resonant integral. Because the two sub-rings are linked end to end,
the 0th atom of a (b) sub-ring is the $N$th atom of b (a) sub-ring. Thus,
the boundary condition of M\"{o}bius molecular ring is given by $a_{0}=b_{N}$ and
$b_{0}=a_{N}$. The Hamiltonian can be rewritten as
\begin{equation}
H=\sum_{j=0}^{N-1}\left[B_{j}^{\dagger}V\sigma_{z}B_{j}-\xi(B_{j}^{\dagger}QB_{j+1}+\textrm{h.c.})\right],
\end{equation}
by a local unitary transformation,
\begin{eqnarray}
B_{j} & \equiv & \left[\begin{array}{c}
c_{j\uparrow}\\
c_{j\downarrow}
\end{array}\right]=U_{j}A_{j},
\end{eqnarray}
\begin{equation}
U_{j}=\frac{1}{\sqrt{2}}\left[\begin{array}{cc}
e^{-i\varphi_{j}/2} & -e^{-i\varphi_{j}/2}\\
1 & 1
\end{array}\right],
\end{equation}
where $c_{j\sigma}$ is the annihilation operator of an electron at
the $j$th nuclear site with $\sigma$ being the pseudo spin label,
$\varphi_{j}=j\delta$, $\delta=2\pi/N,$ and
\begin{align}
Q & =\left[\begin{array}{cc}
e^{i\delta/2} & 0\\
0 & 1
\end{array}\right].
\end{align}
After this unitary transformation, the M\"{o}bius boundary condition can be replaced by the periodical boundary condition, i.e., $B_{N}=B_{0}$.

The Hamiltonian of M\"{o}bius ring can be diagonalized by using the
Fourier transform $B_{j}=\sum_{j=0}^{N-1}e^{-ikj}C_{k}$,
where $C_{k}=\begin{bmatrix}C_{k\uparrow} , C_{k\downarrow}\end{bmatrix}^T$.
We can obtain the two energy sub-bands, i.e.,
\begin{eqnarray}
E_{k\uparrow}&=&V-2\xi\cos\left(k-\delta/2\right),\\
E_{k\downarrow}&=&-V-2\xi\cos k,
\end{eqnarray}
with eigen-states
\begin{eqnarray}
\left|k,\uparrow\right\rangle &=&\frac{1}{\sqrt{2N}}\sum_{j=0}^{N-1}e^{-i(k-\delta/2)j}(a_{j}^{\dagger}-b_{j}^{\dagger})\left|0\right\rangle,\label{eq:stateup}\\
\left|k,\downarrow\right\rangle &=&\frac{1}{\sqrt{2N}}\sum_{j=0}^{N-1}e^{-ikj}(a_{j}^{\dagger}+b_{j}^{\dagger})\left|0\right\rangle,\label{eq:statedown}
\end{eqnarray}
respectively, where $\left|0\right\rangle $ is the state of vacuum,
$k=0,\pm\delta,\pm2\delta\cdots$.
The energy spectrum of M\"{o}bius molecular ring is shown in Fig.~\ref{Figure 1}(b).
Note that the upper band is symmetric with respect to $k=\delta/2$,
while the lower band is symmetric with respect to $k=0$.
Due to this symmetry, the three pairs of intra-band transitions denoted the arrows
with the same color possess the same transition frequencies, respectively.

\subsection{Without Local Field Correction}

In order to judge whether the material is a negative-refraction medium,
we must calculate the relative permittivity $\overleftrightarrow{\varepsilon_{r}}$
and permeability $\overleftrightarrow{\mu_{r}}$ for the same incident frequency. According to Ref.~\cite{Jackson99}, the electric displacement field $\vec{D}$ could be given as
\begin{equation}
\vec{D}=\varepsilon_{0}\overleftrightarrow{\varepsilon_{r}}\vec{E}=\varepsilon_{0}\vec{E}+\vec{P},\label{eq:D}
\end{equation}
where $\varepsilon_{0}$ is the permittivity of vacuum, $\vec{E}$ is the applied electric field, $\vec{P}$ is the polarization field.
And the magnetic induction $\vec{B}$ is
\begin{equation}
\vec{B}=\mu_{0}\left(\vec{H}+\vec{M}\right)=\mu_{0}\overleftrightarrow{\mu_{r}}\vec{H},\label{eq:B}
\end{equation}
where $\mu_{0}$ is the permeability of vacuum, $\vec{H}$ is the applied
magnetic field, $\vec{M}$ is the magnetization field. Under the dipole approximation \cite{Jackson99}, according
to the linear response theory \cite{Kubo85}, we can obtain
\begin{eqnarray}
\vec{P}\!\!\!&=&\!\!\!-\underset{i\neq f}{\sum}\frac{n_{i}\vec{d}_{if}\vec{d}_{fi}\cdot\vec{E}\left(t\right)}{\left(n_{f}+1\right)\hbar v_{0}}\textrm{Re}\left(\frac{1}{\omega-\triangle_{fi}+i\gamma}\right),\label{eq:P}\\
\vec{M}\!\!\!&=&\!\!\!-\underset{i\neq f}{\sum}\frac{n_{i}\mu_{0}\vec{m}_{if}\vec{m}_{fi}\cdot\vec{H}\left(t\right)}{\left(n_{f}+1\right)\hbar v_{0}}\textrm{Re}\left(\frac{1}{\omega-\triangle_{fi}+i\gamma}\right),\label{eq:M}
\end{eqnarray}
where $n_{i}$ and $n_{f}$ are the number of electrons occupying
in the initial and final states respectively, $\triangle_{fi}$ is the transition frequency
between the final state $\left|f\right\rangle $ and the initial state $\left|i\right\rangle$,
$\vec{d}_{if}=\left\langle i\right|\vec{d}\left|f\right\rangle $
and $\vec{m}_{if}=\left\langle i\right|\vec{m}\left|f\right\rangle $
are the matrix elements of electric dipole $\vec{d}$ and magnetic dipole $\vec{m}$ between the initial and final states, $v_0\simeq2\pi(R+W)^{2}W$ is the volume occupied by a M\"{o}bius molecule,
$\omega$ is the frequency of incident light, $\gamma^{-1}$ is the lifetime of the excited states.
Inserting Eq.~(\ref{eq:P}) and Eq.~(\ref{eq:M}) into Eq.~(\ref{eq:D})
and Eq.~(\ref{eq:B}) respectively, the relative permittivity $\overleftrightarrow{\varepsilon_{r}}$
and permeability could be obtained as
\begin{eqnarray}
\overleftrightarrow{\varepsilon_{r}}&=&1-\underset{i\neq f}{\sum}\frac{n_{i}\vec{d}_{if}\vec{d}_{fi}}{\left(n_{f}+1\right)\hbar\varepsilon_{0}v_{0}}\textrm{Re}\left(\frac{1}{\omega-\triangle_{fi}+i\gamma}\right),\label{eq:permittivity}\\
\overleftrightarrow{\mu_{r}}&=&1-\underset{i\neq f}{\sum}\frac{n_{i}\lyxmathsym{\textmu}_{0}\vec{m}_{if}\vec{m}_{fi}}{\left(n_{f}+1\right)\hbar v_{0}}\textrm{Re}\left(\frac{1}{\omega-\triangle_{fi}+i\gamma}\right).\label{eq:permeability}
\end{eqnarray}

Because the size of the molecule is much smaller than the wavelength of the incident light, it is valid to write the interaction Hamiltonian between the molecule and the incident light under dipole approximation \cite{Jackson99}, i.e.,
\begin{equation}
H_{E}=-\vec{d}\cdot\vec{E}\cos\omega t.\label{eq:HE}
\end{equation}
We assume that
\begin{equation}
\left\langle \phi_{js}\right|\vec{r}\left|\phi_{j's'}\right\rangle =\delta_{jj'}\delta_{ss'}\vec{R}_{js},\label{eq:relement}
\end{equation}
where $\left|\phi_{j+}\right\rangle =a_{j}^{\dagger}\left|0\right\rangle $,
$\left|\phi_{j-}\right\rangle =b_{j}^{\dagger}\left|0\right\rangle $,
$\vec{R}_{j+}\left(\vec{R}_{j-}\right)$ is the position of the $j$th
nuclear in a (b) subring which can be given by
\begin{eqnarray}
\vec{R_{j\pm}}=&~&\left(R\pm W\sin\frac{\varphi_{j}}{2}\right)\cos\varphi_{j}\hat{e}_{x}\notag\\
&+&\left(R\pm W\sin\frac{\varphi_{j}}{2}\right)\sin\varphi_{j}\hat{e}_{y}\pm W\cos\frac{\varphi_{j}}{2}\hat{e}_{z},\label{eq:Rj}
\end{eqnarray}
where $\varphi_{j}=j\delta$ is the azimuthal angle of the $j$th nucleus.
By using Eqs.~(\ref{eq:HE})~(\ref{eq:relement}) and~(\ref{eq:Rj}), we can obtain the matrix elements of $H_{E}$ between the eigen-states of $H$ which are written as Eqs.~(\ref{eq:stateup}),~(\ref{eq:statedown}). Here, we only give the matrix elements of intra-band transitions as follows
\begin{eqnarray}
\left\langle k,\uparrow\right|&H_{E}&\left|k\pm\delta,\uparrow\right\rangle =\frac{eR}{2}\left(E^{(x)}\mp iE^{(y)}\right)\cos\omega t,\\
\left\langle k,\downarrow\right|&H_{E}&\left|k\pm\delta,\downarrow\right\rangle =\frac{eR}{2}\left(E^{(x)}\mp iE^{(y)}\right)\cos\omega t,\\
\left\langle k,\sigma\right|&H_{E}&\left|k\pm2\delta,\sigma\right\rangle =0,
\end{eqnarray}
where $\sigma=\downarrow,\uparrow$.
We can summarize the intra-band transition selection
rules for the electric-dipole operator from these matrix elements of $H_{E}$ as
\begin{eqnarray}
\left|k,\sigma\right\rangle \stackrel{x,y}{\rightleftharpoons}\left|k\pm\delta,\sigma\right\rangle.\label{eq:eselect}
\end{eqnarray}

Similarly, we can obtain the matrix elements of the interaction Hamiltonian under dipole approximation \cite{Jackson99}
\begin{equation}
H_{B}=-\vec{m}\cdot\vec{B}\cos\omega t\label{eq:HB}
\end{equation}
between the eigen-states of $H$ which are written as Eqs.~(\ref{eq:stateup}),~(\ref{eq:statedown}). We also only give the matrix elements of intra-band transitions
\begin{eqnarray}
\left\langle k,\uparrow\right|H_{B}\left|k+\delta,\uparrow\right\rangle =&\frac{eW^{2}\xi}{8\hbar}\left[\cos(k-\delta)-\cos(k+\delta)\right]\notag\\
&\left(iB^{(x)}+B^{(y)}-B^{(z)}\right)\cos\omega t,\\
\left\langle k,\uparrow\right|H_{B}\left|k-\delta,\uparrow\right\rangle =&\frac{eW^{2}\xi}{8\hbar}\left[\cos(k-2\delta)-\cos k\right]\notag\\
&\left(iB^{(x)}-B^{(y)}+B^{(z)}\right)\cos\omega t,\\
\left\langle k,\uparrow\right|H_{B}\left|k+2\delta,\uparrow\right\rangle =&\frac{eW^{2}\xi}{8\hbar}\left[\cos k-\cos(k+\delta)\right]\notag\\
&\left(iB^{(x)}+B^{(y)}\right)\cos\omega t,\\
\left\langle k,\downarrow\right|H_{B}\left|k+\delta,\downarrow\right\rangle =&\frac{eW^{2}\xi}{8\hbar}\left[\cos(k-\frac{\delta}{2})-\cos(k+\frac{3\delta}{2})\right]\notag\\
&\left(iB^{(x)}+B^{(y)}-B^{(z)}\right)\cos\omega t,\\
\left\langle k,\downarrow\right|H_{B}\left|k-\delta,\downarrow\right\rangle =&\frac{eW^{2}\xi}{8\hbar}\left[\cos(k-\frac{3\delta}{2})-\cos(k+\frac{\delta}{2})\right]\notag\\
&\left(-iB^{(x)}+B^{(y)}-B^{(z)}\right)\cos\omega t,\\
\left\langle k,\downarrow\right|H_{B}\left|k+2\delta,\downarrow\right\rangle =&\frac{eW^{2}\xi}{8\hbar}\left[\cos(k+\frac{\delta}{2})-\cos(k+\frac{3\delta}{2})\right]\notag\\
&\left(iB^{(x)}+B^{(y)}\right)\cos\omega t.
\end{eqnarray}
Furthermore, the selection rules for the magnetic-dipole operator are
\begin{eqnarray}
\left|k,\sigma\right\rangle \stackrel{x,y,z}{\rightleftharpoons}\left|k\pm\delta,\sigma\right\rangle ,\;\left|k,\sigma\right\rangle \stackrel{x,y}{\rightleftharpoons}\left|k+2\delta,\sigma\right\rangle.\label{eq:mselect}
\end{eqnarray}
Notice that as the matrix elements of $H_B$ for intra-band transitions are proportional to $W^2$ and those for inter-band transitions are $O(RW)$, the magnetic responses for intra-band transitions have been reduced by a factor of $(R/W)^2=(N/\pi)^2$. Hereafter, we will show by numerical simulation the eigen-values of permeability tensor are always positive around the intra-band transitions.

\begin{figure}
\begin{centering}
\includegraphics[width=8cm]{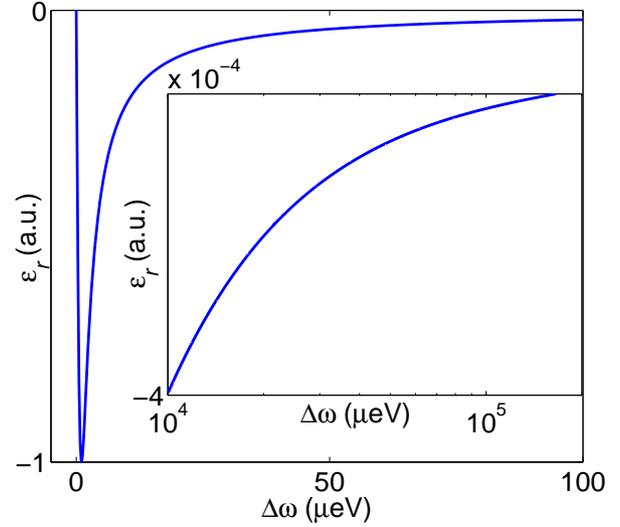}
\par\end{centering}
\caption{The relationship between $\varepsilon_{r}$
and the detuning $\triangle\omega$ around the transition frequency $2.6353829$~eV.
Here we adopt the following parameters: $V=\xi=$3.6~eV \cite{Greenwood1972}, $W=$0.077~nm \cite{Silbey04}, $R=NW/\pi$, $\gamma^{-1}=$4~ns \cite{Tokuji09}.\label{Figure2}}
\end{figure}

According to Eqs.~(\ref{eq:eselect}) and~(\ref{eq:mselect}), only
the transitions $\left|k,\sigma\right\rangle \rightleftharpoons\left|k\pm\delta,\sigma\right\rangle $,
are allowed by both electric and magnetic dipole couplings. Moreover, a transition can take place when the initial state is non-empty (NE), and the final state is not fully filled with electrons (NFF). And we only consider intra-band transitions. Considering all the conditions above, only six transitions, depicted by the arrows in Fig.~\ref{Figure 1}(b),
 are considered in this paper. We divide these transitions into three pairs by the same transition frequencies respectively: (a) $\left|\delta,\uparrow\right\rangle \rightleftharpoons\left|2\delta,\uparrow\right\rangle $, $\left|-\delta,\uparrow\right\rangle \rightleftharpoons\left|-2\delta,\uparrow\right\rangle $, denoted by the red arrows;
(b) $\left|3\delta,\downarrow\right\rangle \rightleftharpoons\left|4\delta,\downarrow\right\rangle $, $\left|-3\delta,\downarrow\right\rangle \rightleftharpoons\left|-4\delta,\downarrow\right\rangle $, denoted by the blue arrows; (c) $\left|4\delta,\downarrow\right\rangle \rightleftharpoons\left|5\delta,\downarrow\right\rangle $, $\left|-4\delta,\downarrow\right\rangle \rightleftharpoons\left|-5\delta,\downarrow\right\rangle $, denoted by the black arrows. We can calculate the elements of
the dielectric tensor by Eq.~(\ref{eq:permittivity}), with the nonvanishing
matrix elements being
\begin{eqnarray}
\varepsilon_{r}^{xx}\!\!\!&=&\!\!\!\varepsilon_{r}^{yy}=1-\sum_{(k\sigma)\in\textrm{NE}}\sum_{(k'\sigma)\in\textrm{NFF}}\eta_{kk'\sigma},\label{eq:epsonxx}\\
\varepsilon_{r}^{yx}\!\!\!&=&\!\!\!-\varepsilon_{r}^{xy} \label{eq:epsonxy}\\
\!\!\!&=&\!\!\!
 \eta_{2\delta,3\delta,\uparrow}^{'}-\eta_{-\delta,-2\delta,\uparrow}^{'}
 +\sum_{k=3\delta,4\delta}\left(\eta_{k,k+\delta,\downarrow}^{'}-\eta_{-k,-k-\delta,\downarrow}^{'}\right)\notag
\end{eqnarray}
where
\begin{equation}
\eta_{kk'\sigma}=\frac{n_{i}e^{2}R^{2}}{4\left(n_{f}+1\right)\hbar\varepsilon_{0}v_{0}}\frac{1}{\omega-\triangle_{kk'\sigma}+i\gamma}\label{eq:eta}
\end{equation}
and $\Delta_{kk'\sigma}$ is the transition frequency between the final
state $\left|k'\sigma\right\rangle $ and the initial state $\left|k\sigma\right\rangle $ within the same band $\sigma$,
$\eta^\prime_{kk'\sigma}$ is
the real part of $\eta_{kk'\sigma}$.
Equation~(\ref{eq:eta}) indicates that if two transitions share the same transition frequency $\Delta_{kk'\sigma}$, they would also possess the same $\eta_{kk'\sigma}$. Since there are three pairs of transitions which possess the same transition frequency, we can obtain three equations
\begin{eqnarray}
\eta_{2\delta,3\delta,\uparrow}&=&\eta_{-\delta,-2\delta,\uparrow},\notag\\
\eta_{3\delta,4\delta,\downarrow}&=&\eta_{-3\delta,-4\delta,\downarrow},\label{eq:etaRelation}\\
\eta_{4\delta,5\delta,\downarrow}&=&\eta_{-4\delta,-5\delta,\downarrow}.\notag
\end{eqnarray}
Inserting these three equations into Eq.~(\ref{eq:epsonxy}), we find that the off-diagonal elements $\varepsilon_{r}^{xy}=\varepsilon_{r}^{yx}=0$. As a result, $\overleftrightarrow{\varepsilon_{r}}$ can be simplified as
\begin{equation}
\overleftrightarrow{\varepsilon_{r}}=\left[\begin{array}{ccc}
1-\eta^\prime & 0 & 0\\
0 & 1-\eta^\prime & 0\\
0 & 0 & 1
\end{array}\right],\label{eq:epsonr}
\end{equation}
where $\eta^{'}=\sum_{(k\sigma)\in\textrm{NE}}\sum_{(k'\sigma)\in\textrm{NFF}}\eta_{kk'\sigma}^{'}$.
The three eigen-values of $\overleftrightarrow{\varepsilon_{r}}$ are
respectively $\varepsilon_{r}^{x}=\varepsilon_{r}^{y}=1-\eta^{'}$
and $\varepsilon_{r}^{z}=1$. Obviously, one of the eigen-values of $\overleftrightarrow{\varepsilon_{r}}$
is identical to 1. Figure~\ref{Figure2} numerically demonstrates the relation between $\varepsilon_{r}^{x}=\varepsilon_{r}^{y}=1-\text{\ensuremath{\eta}}^{'}$
and the detuning $\Delta\omega=\omega-\Delta_{kk^{\prime}\sigma}$.
It presents the situation when
the detuning is less than 100~$\mu$eV, while the inset
shows the relation in the large-detuning regime. We can obtain the bandwidth for negative
permittivity is about 1.5$\times10^{5}~\mu$eV, which is broader than the previous discovery in Ref.~\cite{Cheng2019} by 3 orders of magnitude.

From Eq.~(\ref{eq:eta}), the real part of $\eta_{kk'\sigma}$
could be re-expressed as
\begin{equation}
\eta'_{kk'\sigma}=\frac{n_{i}e^{2}R^{2}}{4\left(n_{f}+1\right)\varepsilon_{0}v_{0}}\left(\Delta\omega_{kk'\sigma}+\frac{\gamma^{2}}{\Delta\omega_{kk'\sigma}}\right)^{-1},\label{eq:etapie}
\end{equation}
where $\Delta\omega_{kk'\sigma}=\omega-\Delta_{kk'\sigma}$.
On account of the initial and final conditions, it can be explicitly written as
\begin{align}
\eta'=  \frac{e^{2}R^{2}}{2\varepsilon_{0}v_{0}}&\left[\left(\Delta\omega_{4\delta,5\delta,\downarrow}+\frac{\gamma^{2}}{\Delta\omega_{4\delta,5\delta,\downarrow}}\right)^{-1}\right.\notag\\
 & +2\left(\Delta\omega_{2\delta,3\delta,\uparrow}+\frac{\gamma^{2}}{\Delta\omega_{2\delta,3\delta,\uparrow}}\right)^{-1}\notag\\
 &\left.+\left(\Delta\omega_{3\delta,4\delta,\downarrow}+\frac{\gamma^{2}}{\Delta\omega_{3\delta,4\delta,\downarrow}}\right)^{-1}\right].\label{eq:etapie1}
\end{align}
To find the bandwidth of negative $\varepsilon_{r}$, we should find the two solutions to the equation
\begin{equation}
1-\eta'\left(\omega\right)=0.\label{eq:1minuetapie}
\end{equation}

For the solution between $\omega_{2\delta,3\delta,\uparrow}=3.2276717$~eV
and $\omega_{4\delta,5\delta,\downarrow}=2.6353829$~eV, which yields $\Delta\omega_{3\delta,4\delta,\downarrow}\gg\Delta\omega_{2\delta,3\delta,\uparrow},\Delta\omega_{4\delta,5\delta,\downarrow}$, Eq.~(\ref{eq:etapie1}) could be simplified as
\begin{align}
\eta'\simeq  \frac{e^{2}R^{2}}{2\varepsilon_{0}v_{0}}&\left[\left(\Delta\omega_{4\delta,5\delta,\downarrow}+\frac{\gamma^{2}}{\Delta\omega_{4\delta,5\delta,\downarrow}}\right)^{-1}\right.\notag\\
 &\left.+2\left(\Delta\omega_{2\delta,3\delta,\uparrow}+\frac{\gamma^{2}}{\Delta\omega_{2\delta,3\delta,\uparrow}}\right)^{-1}\right].\label{eq:etapie2}
\end{align}
For the present parameters, we find $\gamma\sim10^{-6}$~eV and $C=\frac{e^{2}R^{2}}{2\varepsilon_{0}v_{0}}\sim11.4$~eV.
When $\Delta\omega_{2\delta,3\delta,\uparrow}$ and $\Delta\omega_{4\delta,5\delta,\downarrow}\gg\gamma$
is satisfied, the terms of $\gamma^{2}$ could be ignored, and thus
\begin{equation}
\eta'\simeq
C\left(\frac{1}{\Delta\omega_{4\delta,5\delta,\downarrow}}+\frac{2}{\Delta\omega_{2\delta,3\delta,\uparrow}}\right).
\end{equation}
Inserting the above equation into Eq.~(\ref{eq:1minuetapie}), we obtain the solution $\omega_{1}=2.8305463$~eV while the other solution $\omega=37.23251$~eV should be discarded because it is far away from the transition.
To find the solution around the resonance frequency $\omega=2.6353829$~eV, we simplify $\eta'$ as
\begin{align}
\eta'=  C\left(\Delta\omega_{4\delta,5\delta,\downarrow}+\frac{\gamma^{2}}{\Delta\omega_{4\delta,5\delta,\downarrow}}\right)^{-1}.\label{eq:etapie3}
\end{align}
Inserting Eq.~(\ref{eq:etapie3}) into Eq.~(\ref{eq:1minuetapie}) yields $\Delta\omega_{4\delta,5\delta,\downarrow}\simeq 0$ as $C^{2}\gg\gamma^{2}$. The other solution $\Delta\omega_{4\delta,5\delta,\downarrow}\simeq C$ should be discarded because it is not close to the transition. Thus, the other solution to Eq.~(\ref{eq:1minuetapie}) is $\omega_{2}=2.6353829$~eV.
We obtain the window of negative permittivity is $\Delta\omega=\omega_{1}-\omega_{2}=0.1952$~eV,
which is consistent with the numerical simulation in Fig.~\ref{Figure2}.

\begin{figure}
\begin{centering}
\includegraphics[width=8cm]{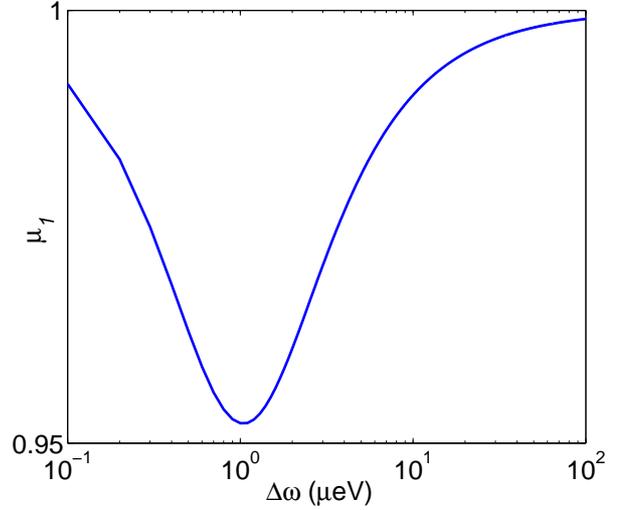}
\par\end{centering}
\caption{The relationship between $\mu_{1}$
and the detuning $\triangle\omega$ around the transition frequency $3.2276717$~eV.
Here we adopt the following parameters: $V=\xi=$3.6~eV \cite{Greenwood1972}, $W=$0.077~nm \cite{Silbey04}, $R=NW/\pi$, $\gamma^{-1}=$4~ns \cite{Tokuji09}.\label{Figure3}}
\end{figure}

In the same way, we can calculate the elements of the permeability
tensor $\overleftrightarrow{\mu_{r}}$ by using Eq.~(\ref{eq:permeability}) as
\begin{equation}
\overleftrightarrow{\mu_{r}}=\begin{pmatrix}1-\beta & -i\beta_{1} & i\beta_{1}\\
i\beta_{1} & 1-\beta & \beta\\
-i\beta_{1} & \beta & 1-\beta
\end{pmatrix},
\end{equation}
where $\beta=\sum_{(k\sigma)\in\textrm{NE}}\sum_{(k'\sigma)\in\textrm{NFF}}\alpha_{kk'\sigma}^{2}\text{\ensuremath{\eta}}_{kk'\sigma}^{'},$
and
\begin{align}
\beta_{1}&=  \alpha_{2\delta,3\delta,\uparrow}^{2}\eta_{2\delta,3\delta,\uparrow}^{'}-\alpha_{-\delta,-2\delta,\uparrow}^{2}\eta_{-\delta,-2\delta,\uparrow}^{'}\label{eq:beita1}\\
 &+\sum_{k=3\delta,4\delta}\left(\alpha_{k,k+\delta',\downarrow}^{2}\eta_{k,k+\delta,\downarrow}^{'}-\alpha_{-k,-k-\delta',\sigma}^{2}\eta_{-k,-k-\delta,\downarrow}^{'}\right),\notag
\end{align}
\begin{align}
&\alpha_{k,k+\delta,\uparrow}=\frac{W^{2}\xi}{4\hbar cR}\left[\cos(k-\delta)-\cos(k+\delta)\right],\notag\\
&\alpha_{k,k-\delta,\uparrow}=\frac{W^{2}\xi}{4\hbar cR}\left[\cos(k-2\delta)-\cos(k)\right],\label{eq:alpha}\\
&\alpha_{k,k+\delta,\downarrow}=\frac{W^{2}\xi}{4\hbar cR}\left[\cos\left(k-\frac{\delta}{2}\right)-\cos\left(k+\frac{3\delta}{2}\right)\right],\notag\\
&\alpha_{k,k-\delta,\downarrow}=\frac{W^{2}\xi}{4\hbar cR}\left[\cos\left(k-\frac{3\delta}{2}\right)-\cos\left(k+\frac{\delta}{2}\right)\right].\notag
\end{align}

According to Eq.~(\ref{eq:alpha}), we can obtain the following three relations,
\begin{eqnarray}
\alpha_{2\delta,3\delta,\uparrow}^{2}=\alpha_{-\delta,-2\delta,\uparrow}^{2},\notag\\
\alpha_{3\delta,4\delta,\downarrow}^{2}=\alpha_{-3\delta,-4\delta,\downarrow}^{2},\label{eq:alphaRelation}\\
\alpha_{4\delta,5\delta,\downarrow}^{2}=\alpha_{-4\delta,-5\delta,\downarrow}^{2}.\notag
\end{eqnarray}
Inserting Eqs.~(\ref{eq:alphaRelation}) and~(\ref{eq:etaRelation})
into Eq.~(\ref{eq:beita1}), we can obtain $\beta_{1}=0$.
Therefore, $\overleftrightarrow{\mu_{r}}$ could be simplified as
\begin{equation}
\overleftrightarrow{\mu_{r}}=\begin{pmatrix}1-\beta & 0 & 0\\
0 & 1-\beta & \beta\\
0 & \beta & 1-\beta
\end{pmatrix}.\label{eq:mur}
\end{equation}
The permeability tensor is not diagonal in the molecular coordinate
system as shown above. As a result, $\overleftrightarrow{\varepsilon_{r}}$
and $\overleftrightarrow{\mu_{r}}$ cannot be simultaneously diagonalized by the
same rotation transformation.

The permeability possess three eigen-values, i.e., $\mu_1=1-2\beta$, $\mu_2=1-\beta$ and $\mu_3=1$. Because $\alpha_{kk'\sigma}\sim10^{-3}$, $\mu_j$'s are generally less significantly influenced by the medium than $\varepsilon_r^{j}$'s ($j=,x,y,z$).
This prediction is numerically confirmed in Fig.~\ref{Figure3}. As shown,
all eigen-values of $\mu_r$ are positive.

\subsection{Local Field Correction}

In the above sections, we obtain the relative permittivity and permeability by the linear response theory. However, because all molecules in the medium are polarized by the applied fields, the total field experienced by a molecule is the sum of the external field $\vec{E}$ and internal field $\vec{E}_{i}$ \cite{Jackson99}, i.e.,
\begin{equation}
\vec{E}_\mathrm{tot}=\vec{E}+\vec{E}_{i}.
\end{equation}
And internal field could be written as $\vec{E}_{i}=\vec{E}_{\mathrm{near}}-\vec{E}_{\mathrm{mean}}$,
where $\vec{E}_{\mathrm{near}}$ is the electric field produced
by nearby molecules and $\vec{E}_{\mathrm{mean}}$ is the mean field, which is evaluated as
\begin{equation}
\vec{E}_{\mathrm{mean}}=-\frac{1}{3\varepsilon_{0}}\sum_{l}\frac{\vec{p}_{l}}{V},
\end{equation}
where $\vec{p}_{l}$ is the induced dipole moment
of the $l$th molecule inside the volume $V$. For a sufficiently-weak
field, the induced dipole moment is given by
\begin{equation}
\vec{p}_{l}=\varepsilon_{0}\gamma_\textrm{mol}\vec{E}_\textrm{tot},\label{eq:pl}
\end{equation}
where $\gamma_\textrm{mol}$ is the molecular polarizability. According to linear response theory \cite{Kubo85}, the electric dipole is written as
\begin{equation}
\left\langle \vec{d}\right\rangle =-\sum_{i\neq f}\frac{\vec{d}_{if}\vec{d}_{fi}\cdot\vec{E}_\textrm{tot}}{\hbar}\textrm{Re}\left(\frac{1}{\omega-\triangle_{fi}+i\gamma}\right).\label{eq:d}
\end{equation}
Because $\vec{p}_{l}=\left\langle \vec{d}\right\rangle$, due to Eqs.~(\ref{eq:d}) and~(\ref{eq:pl}), we can obtain
\begin{equation}
\gamma_\textrm{mol}=-\underset{i\neq f}{\sum}\frac{\vec{d}_{if}\vec{d}_{fi}}{\hbar\varepsilon_{0}}\textrm{Re}\left(\frac{1}{\omega-\triangle_{fi}+i\gamma}\right).\label{eq:gamma-mol}
\end{equation}
The polarization $\vec{P}=\sum_{l}\vec{p}_{l}/V$ could be written as
\begin{equation}
\vec{P}=\frac{\vec{p}_{l}}{\upsilon_{0}},\label{eq:P-1}
\end{equation}
if we assume identical contributions from all molecules.
And the relationship between $\vec{P}$ and the electric
field is
\begin{equation}
\vec{P}=\varepsilon_{0}\overleftrightarrow{\chi_{e}}\vec{E}.\label{eq:PE}
\end{equation}
By combining Eqs.~(\ref{eq:pl}), (\ref{eq:P-1}) and (\ref{eq:PE}),
we have
\begin{equation}
\overleftrightarrow{\chi_{e}}=\left(1-\frac{\gamma_\textrm{mol}}{3\upsilon_{0}}\right)^{-1}\frac{\gamma_\textrm{mol}}{\upsilon_{0}}.\label{eq:xe}
\end{equation}
Inserting Eq.~(\ref{eq:D}) into Eq.~(\ref{eq:PE}), we obtain
\begin{equation}
\overleftrightarrow{\varepsilon_{r}}=1+\overleftrightarrow{\chi_{e}}.\label{eq:xe-epson}
\end{equation}
Inserting Eq.~(\ref{eq:xe}) into Eq.~(\ref{eq:xe-epson}), $\overleftrightarrow{\varepsilon_{r}}$
could be expressed in terms of $\gamma_\textrm{mol}$ as
\begin{equation}
\overleftrightarrow{\varepsilon_{r}}=1+\left(1-\frac{\gamma_\textrm{mol}}{3\upsilon_{0}}\right)^{-1}\frac{\gamma_\textrm{mol}}{\upsilon_{0}}.
\end{equation}
Inserting Eq.~(\ref{eq:gamma-mol}) into the above equation, we can
obtain $\overleftrightarrow{\varepsilon_{r}}$ tensor with the nonvanishing
matrix elements
\begin{eqnarray}
\varepsilon_{r}^{xx}&=&\varepsilon_{r}^{yy}=\frac{3-2\eta'}{3+\eta'},\label{eq:epsonr-local}\label{eq:epsonlocal}\\
\varepsilon_{r}^{zz}&=&1.
\end{eqnarray}

\begin{figure}
\begin{centering}
\includegraphics[width=8cm]{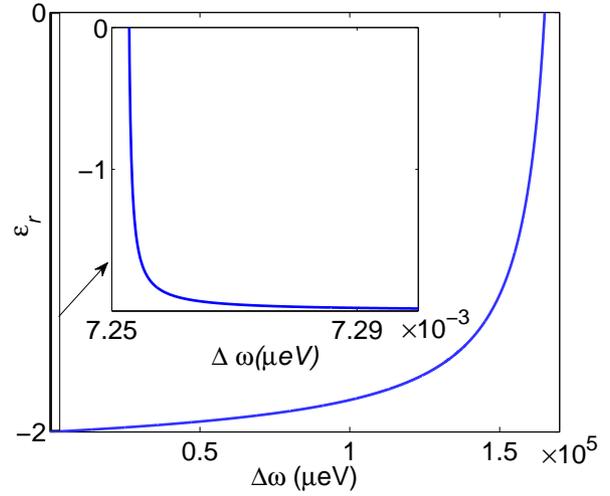}
\par\end{centering}
\caption{$\varepsilon_{r}^{xx}$ ($\varepsilon_{r}^{yy}$) modified by local field effect. \label{Figure4}}
\end{figure}

It follows from Eq.~(\ref{eq:epsonlocal}) that the bandwidth of negative
permittivity, when we consider the local field effect, is determined
by the solution to
\begin{equation}
\eta'=\frac{3}{2}.\label{eq:etapie32}
\end{equation}
Comparing Eq.~(\ref{eq:etapie32}) to Eq.~(\ref{eq:1minuetapie}),
we find that the bandwidth of negative permittivity is modified by local
field effect only with a factor $3/2$. Base on Eq.~(\ref{eq:etapie32}),
we present the relation between the relative permittivity modified by local field effect
and the detuning in Fig.~\ref{Figure4}. Comparing Fig.~\ref{Figure2}
to Fig.~\ref{Figure4}, we find that local field effect only slightly changes the bandwidth of negative permittivity.
In the same way, we can find that local field effect only slightly changes the permeability, and the three eigen-values of permeability tensor are all positive.

\section{Negative Refraction with Linearly-Polarized Incident Light}

In the previous section, we have calculated the relative permittivity and permeability of the M\"{o}bius medium. The relative permittivity and permeability are second-order tensors, which are not diagonal in the same coordinate system. In this section, by both analytic and numerical simulations, we clearly show that there is hyperbolic dispersion relation in the M\"{o}bius medium, and the conditions under which the negative refraction can take place is discussed.

\subsection{$H$-Polarized Incident Configuration}

\begin{figure}
\begin{centering}
\includegraphics[width=8cm]{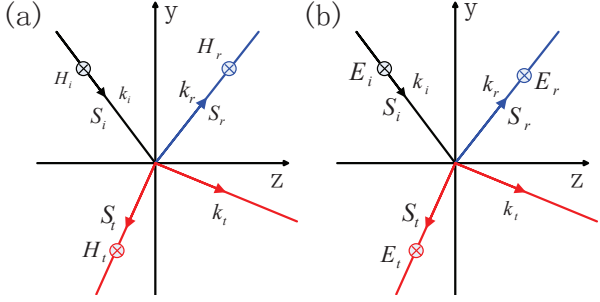}
\par\end{centering}
\caption{\label{Figure5}Negative refraction for hyperbolic dispersion with
$\varepsilon_{r}^{x}<0$, $\varepsilon_{r}^{y}<0$, and $\varepsilon_{r}^{z}>0$:
(a) $H$-polarized incident field; (b) $E$-polarized incident field.
The Poynting vector, wave vector and magnetic field of the incident
wave are $\vec{S}_{i}$, $\vec{k}_{i}$
and $\vec{H}_{i}$, respectively. It is reflected
with Poynting vector $\vec{S}_{r}$, wave vector
$\vec{k}_{r}$ and magnetic field $\vec{H}_{r}$.
It is transmitted with Poynting vector $\vec{S}_{t}$,
wave vector $\vec{k}_{t}$ and magnetic field
$\vec{H}_{t}$. }
\end{figure}
 As illustrated in Fig.~\ref{Figure5}(a), a linearly-polarized monochromatic light is incident
from the air into M\"{o}bius medium. The electric and magnetic fields of incident wave are respectively
\begin{eqnarray}
\vec{E}_{i}&=&\left(E_{iy}\hat{e}_{y}+E_{iz}\hat{e}_{z}\right)e^{i\left(\vec{k}_{i}\cdot\vec{r}-\omega t\right)},\\
\vec{H}_{i}&=&H_{ix}\hat{e}_{x}e^{i\left(\vec{k}_{i}\cdot\vec{r}-\omega t\right)},\label{eq:Hi}
\end{eqnarray}
where $\vec{k}_{i}=k_{iy}\hat{e}_{y}+k_{iz}\hat{e}_{z}$ is the wave
vector. We call this incident configuration as $H$-polarized light
because the magnetic field of the incident light is perpendicular
to the wave vector.

According to the boundary conditions,
\begin{eqnarray}
\hat{e}_{y}&\times&\left(\vec{H}_{i}+\vec{H}_{r}-\vec{H}_{t}\right)=0,\\
k_{tx}&=&k_{ix}=0,\label{eq:boundary-H}\\
k_{tz}&=&k_{iz}>0,
\end{eqnarray}
and Eq.~(\ref{eq:Hi}),
the magnetic and electric fields of the refracted light can be written as
\begin{eqnarray}
\vec{H}_{t}&=&H_{tx}\hat{e}_{x}e^{i\left(\vec{k}_{t}\cdot\vec{r}-\omega t\right)},\label{eq:Ht}\\
\vec{E}_{t}&=&\left(E_{ty}\hat{e}_{y}+E_{tz}\hat{e}_{z}\right)\mathrm{e^{\mathit{i\left(\vec{k}_{t}\cdot\vec{r}-\omega t\right)}}},\label{eq:Et}
\end{eqnarray}
where the wave vector of refracted wave is $\vec{k}_{t}=k_{ty}\hat{e}_{y}+k_{tz}\hat{e}_{z}$.
According to Eqs.~(\ref{eq:Ht}) and~(\ref{eq:Et}), the Maxwell's equations of refracted light could be written as
\begin{eqnarray}
\vec{\nabla}\times\vec{E}_{t}&=&i\omega\mu_{0}\overleftrightarrow{\mu_{r}}\vec{H}_{t},\label{eq:po-ke}\\
\vec{\nabla}\times\vec{H}_{t}&=&-i\omega\varepsilon_{0}\overleftrightarrow{\varepsilon_{r}}\vec{E}_{t}.\label{eq:po-kh}
\end{eqnarray}
Inserting Eq.~(\ref{eq:po-kh}) into Eq.~(\ref{eq:po-ke}), we can obtain
\begin{equation}
\vec{\nabla}\times\left[\left(\overleftrightarrow{\varepsilon_{r}}\right)^{-1}\vec{\nabla}\times\vec{H}_{t}\right]=\frac{\omega^{2}}{c^{2}}\overleftrightarrow{\mu_{r}}\vec{H}_{t}.\label{eq:matrixH}
\end{equation}
For nontrivial solutions to the equation, the
determinant of it's coefficient matrix should be equal to zero, which yields a hyperbolic dispersion relation
\begin{equation}
\varepsilon_{r}^{y}k_{ty}^{2}+\varepsilon_{r}^{z}k_{tz}^{2}=\frac{\omega^{2}}{c^{2}}\varepsilon_{r}^{y}\varepsilon_{r}^{z}\mu_{r}^{xx},\label{eq:dispersion-1}
\end{equation}
where the solutions are
\begin{equation}
k_{ty}=\pm\sqrt{\omega^{2}\varepsilon_{r}^{z}\mu_{r}^{xx}/c^{2}-\varepsilon_{r}^{z}k_{tz}^{2}/\varepsilon_{r}^{y}}.\label{eq:solution-kty}
\end{equation}
Because $\varepsilon_{r}^{z}=1$ and $\varepsilon_{r}^{y}<0$,
the real solution to Eq.~(\ref{eq:solution-kty}) always exists. Below, we will show
that we should choose the negative solution for a correct Poynting vector
of the refracted light. According to Eq.~(\ref{eq:po-kh}),
we can obtain
\begin{equation}
\vec{E}_{t}=-\frac{1}{\omega\varepsilon_{0}}\left(\overleftrightarrow{\varepsilon_{r}}\right)^{-1}\left(\vec{k}_{t}\times\vec{H}_{t}\right).
\end{equation}
Inserting the above equation into $\vec{S}_{t}=\frac{1}{2}\left(\vec{E}_{t}\times\vec{H}_{t}^{*}\right)$,
the Poynting vector of refracted light could be written as $\vec{S}_{t}=S_{ty}\hat{e}_{y}+S_{tz}\hat{e}_{z}$
with
\begin{eqnarray}
S_{ty}&=&\frac{k_{ty}H_{tx}^{2}}{2\omega\varepsilon_{0}\varepsilon_{r}^{z}},\label{eq:Sty}\\
S_{tz}&=&\frac{k_{tz}H_{tx}^{2}}{2\omega\varepsilon_{0}\varepsilon_{r}^{y}}.\label{eq:Stz}
\end{eqnarray}
As shown in Fig.~\ref{Figure5}(a), the condition under which the refracted
light can propagate in the medium is $S_{ty}<0$. Because $\varepsilon_{r}^{z}=1>0$, according to Eq.~(\ref{eq:Sty}), we should take the negative solution in Eq.~(\ref{eq:solution-kty}) to meet the criterion $S_{ty}<0$.
According to the boundary conditions in Eq.~(\ref{eq:boundary-H}), we can obtain from Eq.~(\ref{eq:Stz}) that $S_{tz}<0$ as $\varepsilon_{r}^{y}<0$.
Because the Poynting vectors of incident and refracted lights are on
the same side of the normal, negative refraction is realized. The bandwidth of negative refraction is given by the bandwidth of negative $\varepsilon_{r}^{y}$, i.e., 0.1952~eV.

\subsection{$E$-Polarized Incident Configuration}

\begin{figure}
\begin{centering}
\includegraphics[width=9cm]{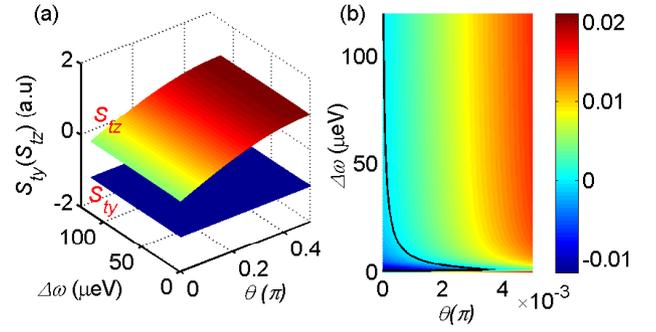}
\par\end{centering}
\caption{(a)\label{Figure6} The dependence of $S_{ty}$ and $S_{tz}$
on $\Delta\omega$ and $\theta$. (b) The bandwidth of negative $S_{tz}$ vs $\theta$.
The black line is the contour line for $S_{tz}=0$, which separates the regime $S_{tz}<0$ at the bottom-left and the regime $S_{tz}>0$ at
the top-right.}
\end{figure}

Analogously, we consider an $E$-polarized incident configuration, i.e.,
\begin{eqnarray}
\vec{E}_{i}&=&E_{ix}\hat{e}_{x}\mathrm{e}^{i\left(\vec{k}_{i}\cdot\vec{r}-\omega t\right)},\\
\vec{H}_{i}&=&\left(H_{iy}\hat{e}_{y}+H_{iz}\hat{e}_{z}\right)\mathrm{e}^{i\left(\vec{k}_{i}\cdot\vec{r}-\omega t\right)},
\end{eqnarray}
where $\vec{k}_{i}=k_{iy}\hat{e}_{y}+k_{iz}\hat{e}_{z}$, $\vec{E}_{i}$ and $\vec{H}_{i}$ are the electric and magnetic fields of incident wave, respectively.
In a similar way of obtaining Eq.~(\ref{eq:matrixH}), we can obtain
the equation
\begin{equation}
\vec{\nabla}\times\left[\left(\overleftrightarrow{\mu_{r}}\right)^{-1}\vec{\nabla}\times\vec{E}_{t}\right]=\frac{\omega^{2}}{c^{2}}\overleftrightarrow{\varepsilon_{r}}\vec{E}_{t}.\label{eq:matrixE}
\end{equation}
The requirement for nontrivial solution yields
the hyperbolic dispersion relation as
\begin{equation}
\varepsilon_{r}^{y}k_{ty}^{2}+\varepsilon_{r}^{z}k_{tz}^{2}=\frac{\omega^{2}}{c^{2}}\varepsilon_{r}^{y}\varepsilon_{r}^{z}\mu_{r}^{xx},\label{eq:dispersion-2}
\end{equation}
which is the same as Eq.~(\ref{eq:dispersion-1}) with
the same solution given by Eq.~(\ref{eq:solution-kty}). We can obtain
the Poynting vector of refracted light $\vec{S}_{t}=S_{ty}\hat{e}_{y}+S_{tz}\hat{e}_{z}$,
with
\begin{eqnarray}
S_{ty}&=&\frac{E_{tx}^{2}}{2\omega\mu_{0}\mu_{1}}\left(\mu_{r}^{yz}k_{tz}+\mu_{r}^{xx}k_{ty}\right),\label{eq:Sty-E}\\
S_{tz}&=&\frac{E_{tx}^{2}}{2\omega\mu_{0}\mu_{1}}\left(\mu_{r}^{xx}k_{tz}+\mu_{r}^{yz}k_{ty}\right).\label{eq:Stz-E}
\end{eqnarray}
Figure~\ref{Figure3} illustrates that $\mu_{1}>0$. As shown in Fig.~\ref{Figure5}(b),
$S_{ty}$ must be negative, otherwise there would be no refracted light.
By numerical calculation, we find that we should choose the negative sign
of $k_{ty}$ in Eq.~(\ref{eq:solution-kty}) to ensure $S_{ty}<0$.
Figure~\ref{Figure5}(b) shows that the conditions of negative refraction
are $S_{ty}<0$ and $S_{tz}<0$. Because the coefficient $E_{tx}^{2}/\left(2\omega\mu_{0}\mu_{1}\right)>0$,
the conditions could be written as $\mu_{r}^{yz}k_{tz}+\mu_{r}^{xx}k_{ty}<0$
and $\mu_{r}^{xx}k_{tz}+\mu_{r}^{yz}k_{ty}<0$. Inserting the boundary
conditions
\begin{equation}
k_{tx}=k_{ix}=0,\quad k_{tz}=k_{iz}=k_{i}\sin\theta,
\end{equation}
where $\theta$ is the angle of incidence, and Eq.~(\ref{eq:mur})
into Eqs.~(\ref{eq:Sty-E}) and (\ref{eq:Stz-E}), we plot Fig.~\ref{Figure6}(a),
which shows $S_{ty}<0$ and $S_{tz}$ can change sign along with $\theta$ and $\Delta\omega$. We further plot $S_{tz}$ vs $\theta$ and $\Delta\omega$ in Fig.~\ref{Figure6}(b).
As shown, when the incident angle $\theta$ is enlarged,
the bandwidth of negative refraction is narrowed.
Generally, the bandwidth for $H$-polarized incident configuration is much wider than that for $E$-polarized incident configuration.

\section{Conclusion}

In this work, we propose a new approach to realize negative refraction in chiral molecules by using hyperbolic dispersion. When we consider intra-band transitions, all of the
three eigen-values of $\overleftrightarrow{\mu_{r}}$ are positive
for the whole frequency domain, and one of the three eigen-values
of $\overleftrightarrow{\varepsilon_{r}}$ possesses a different sign from the remain two in some frequency domains. The window of negative refraction is determined
by the window of negative $\overleftrightarrow{\varepsilon_{r}}$.
Since the electric response is generally larger than the magnetic response orders of magnitude, the hyperbolic metamaterial
can significantly broaden the window of negative refraction.
In M\"{o}bius medium, since the transition frequencies of intra-band transition are smaller than those of the inter-band transitions, we can observe a bandwidth with 0.1952~eV around $\omega=$2.6354~eV ($471.4$~nm), which is in the range of visible light. Compared to the previous proposals in Refs.~\cite{Fang2016,Cheng2019}, the bandwidth of negative refraction has been significantly broadened by 3 orders of magnitude and the center frequency has been shifted from the ultraviolet to the visible frequency domain.

This work was supported by National Natural Science Foundation of China under the grants Nos.~11505007, 11674033, 11474026.

\end{document}